\documentclass[aps,twocolumn,prc,floatfix,showpacs,preprintnumbers,amsmath,amssymb,nofootinbib,groupedaddress]{revtex4-1}
\usepackage[normalem]{ulem}
\usepackage{wasysym}
\usepackage{color}
\usepackage{graphicx}
\usepackage{dcolumn}    
\usepackage{bigstrut}
\usepackage{lipsum}
\begin{document}
%
\title{The nuclear symmetry energy and the breaking of the isospin symmetry: \\ how do they reconcile with each other?}

\author{X. Roca-Maza}
\email{xavier.roca.maza@mi.infn.it}
\affiliation{Dipartimento di Fisica, Universit\`a degli Studi di Milano and INFN,  
Sezione di Milano, Via Celoria 16, 20133 Milano, Italy.}

\author{G.  Col\`o}
\email{colo@mi.infn.it}
\affiliation{Dipartimento di Fisica, Universit\`a degli Studi di Milano and INFN,  
Sezione di Milano, Via Celoria 16, 20133 Milano, Italy.}

\author{H. Sagawa}
\email{hiroyuki.sagawa@gmail.com}
\affiliation{RIKEN Nishina Center, Wako 351-0198, Japan\\ 
Center for Mathematics and Physics, University of Aizu, Aizu-Wakamatsu, Fukushima 965-8560, Japan}

\date{\today}

\begin{abstract}
We analyze and propose a solution to the apparent inconsistency between 
our current knowledge of the Equation of State of asymmetric nuclear 
matter, the energy of the Isobaric Analog State (IAS) in a heavy nucleus 
such as ${}^{208}$Pb, and the isospin symmetry breaking forces in the nuclear medium. 
This is achieved by performing state-of-the-art Hartree-Fock plus Random Phase Approximation 
calculations of the IAS 
that include all isospin symmetry breaking contributions. To this aim, we propose a new effective interaction 
that is successful in reproducing the IAS excitation energy without compromising other properties of finite nuclei. 
\end{abstract}


\maketitle

The nuclear physics community has been striving for quite some time to determine the symmetry energy, 
and in particular its density dependence \cite{topical_volume}. The symmetry energy is the energy per 
particle needed to change protons into neutrons in uniform matter at a given density $\rho$. 
At saturation density of symmetric matter, $\rho_0\approx$ 0.16 fm$^{-3}$, its value is 
between 29-32.7 MeV \cite{Lattimer:2013} or between 30.7-32.5 MeV \cite{Li:2013} if one performs 
a weighted average of various extractions, but a broader interval, namely 28.5-34.5 MeV, has been extracted 
in Ref. \cite{oertel2017} (cf. also Ref. \cite{Horowitz:2014}). 
In short, we still do not know precisely the value of the symmetry energy at saturation density and, as we argue
below,
its density dependence 
is even more uncertain.

A deeper understanding would be highly needed, because the accurate characterization of the symmetry 
energy entails profound consequences for the study of the neutron distributions in nuclei along 
the whole nuclear chart, as well of other properties of neutron-rich nuclei \cite{topical_volume}. Its knowledge impacts on heavy-ion reactions where the neutron-proton 
imbalance varies between the incoming and outgoing interacting nuclei \cite{Tsang2011}. The symmetry 
energy is also of paramount importance for understanding the properties of compact objects like neutron 
stars: it directly impacts, for instance, the determination of the radius of a low-mass neutron 
star \cite{Carriere2003}, and is also crucial for understanding stars with a larger mass where the physics 
of nuclear matter above saturation density also enters. Neutron star physics have received a new strong 
boost very recently, as the LIGO-Virgo collaboration announced the first detection of gravitational 
waves from a binary neutron star merger, setting a new type of constraint on the radius of a neutron star \cite{LIGO2017}. Neutron star mergers are also a promising 
site for the r-process nucleosynthesis \cite{Kasen2017}, in which the symmetry energy plays again a substantial role, 
since the r-process path is governed by the mass of neutron-rich nuclei as well as by their beta-decays. 
Last but not least, the knowledge of the nuclear symmetry energy is relevant for Standard Model tests 
via atomic parity violation, as shown, e.g., in Ref. \cite{Sil2005}.

If $\beta$ is the local neutron-proton asymmetry, $\beta \equiv (\rho_n-\rho_p) / \rho$, 
the energy per particle in matter having 
neutron-proton imbalance is a function $\frac{E}{A}(\rho,\beta)$. 
Such function can be expanded in even powers of $\beta$ owing to isospin symmetry (the Coulomb force has to be taken out when dealing with a uniform system). By retaining only the quadratic term we can write
\begin{equation}\label{def_sym}
\frac{E}{A}(\rho,\beta) = \frac{E}{A}(\rho,\beta=0) + S(\rho) \beta^2.
\end{equation}
This equation defines the symmetry energy $S(\rho)$, that is, the difference between the energy per particle $E/A$ in neutron and symmetric matter. Eq. (\ref{def_sym}) clearly explains why an accurate knowledge of the symmetry energy is mandatory in order to establish a link between the physics of finite nuclei and that of a neutron star.

The symmetry energy can be expanded around saturation density as $S(\rho)=J+L\left(\frac{\rho -\rho_0}{\rho_0}\right)+\frac{1}{2}K_{sym}\left(\frac{\rho-\rho_0}{\rho_0}\right)^2+\ldots$, where different parameters have been defined, namely $J\equiv  S(\rho_0)$, $L  \equiv  3\rho_0\ S^\prime(\rho_0)$, and $K_{\rm sym} \equiv 9\rho_0^2\ S^{\prime\prime}(\rho_0)$. On these parameters much attention has been focused. While $K_{\rm sym}$ is basically not known, the error on $L$, 
referred to as the ``slope parameter'', is believed to be still significantly larger than the error on $J$: 
ranges between $\approx$ 40-75 MeV or between $\approx$ 30-90 MeV 
are mentioned in Refs. \cite{Lattimer:2013,Li:2013,Horowitz:2014,oertel2017}.
Many authors have pointed out a correlation between $L$ and the neutron skin $\Delta R_{\rm np}
\equiv \langle r_n^2 \rangle^{1/2} - \langle r_p^2 \rangle^{1/2}
$ of a heavy nucleus like $^{208}$Pb \cite{Brown:2000,Brown:2001,Furnstahl:2002,centelles2009}. This can be understood also in an intuitive way. The larger is the change in symmetry energy as a function of the density, the more convenient the system finds it to push the excess neutrons to a lower density region, that is, towards the surface. Precise and model-independent measurements of the neutron skin are of paramount importance to pin down the value of $L$ \cite{RocaMaza:2011,Brown:2017}. Hence, the relevance of experiments like PREX and possible steps forward in the same direction \cite{prex,horowitz2012,crex}.

The difficulties in determining the symmetry energy behavior are associated with our still incomplete understanding of the strong interaction in the low-energy regime that is important for nuclei. Then, finding a connection with an observable that is {\it not} sensitive to the strong force becomes an asset. The Isobaric Analog State (IAS) is one of the well established properties of nuclei that is measured accurately, and is only sensitive to the isospin symmetry breaking (ISB) in the nuclear medium  due to Coulomb interaction and, to a lesser extent, the other effects that we will discuss below.  Then, here comes the focus of our work. If there is an inconsistency between the properties of the symmetry energy and our knowledge of the IAS and of the ISB forces, this is a serious issue. As discussed above, the neutron skin is strongly correlated  with the density dependence of the symmetry energy. Therefore, we cannot accept that the values of the neutron skin do not match our understanding of the isospin symmetry, one of the basic symmetries of nature, and of its breaking. In this paper, a solution is proposed.

Energy Density Functionals (EDFs) constitute, at present, the only theoretical framework in which the neutron skins and the IAS energies can be consistently calculated from a microscopic perspective, in medium-heavy nuclei \cite{Bender}. Many different parameter sets
exist, for every class of EDFs; basically, there are three classes
of widely used functionals, namely Skyrme, Gogny and relativistic
mean-field (RMF) functionals. We can focus our attention on
some recent, accurate functionals, and in particular on those
that have been used in recent years to correlate the symmetry energy
parameters with some observables. 

\begin{figure}[t!]
\includegraphics[width=0.99\linewidth,clip=true]{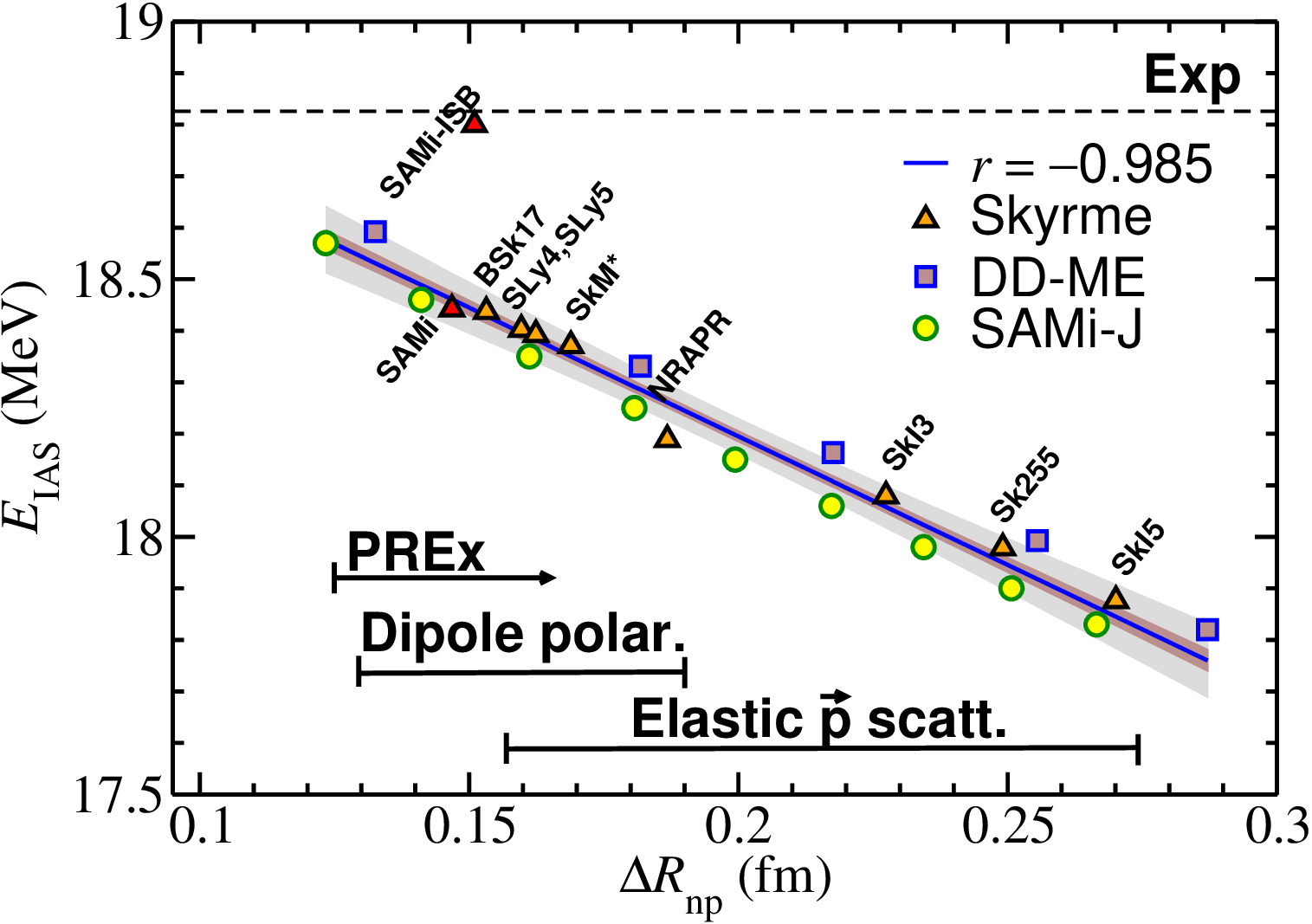}
\caption{Energy of the IAS as a function of $\Delta R_{np}$. The arrows indicate the experimental results from polarized proton elastic scattering \cite{zenihiro2010}, parity violating elastic electron scattering \cite{horowitz2012}, and from the electric dipole polarizability \cite{roca-maza2015}. See the text for a discussion.}
\label{fig1}
\end{figure}

Within the Skyrme functionals, SAMi \cite{RocaMaza2012} has been shown to be specially accurate in the description of charge-exchange resonances such as the Gamow-Teller resonance. Starting from the prototype SAMi functional, a systematically varied family has been generated, by keeping a similar quality of the original fit and varying the values of $J$ and $L$, in Ref. \cite{RocaMaza2013}. In addition, a family based also on the systematic variation of $J$ and $L$ with respect to a RMF with density dependent meson-nucleon vertices (DD-ME \cite{ddme}), was also introduced. These functionals provide values of the neutron skin through the Hartree-Fock (HF) or Hartree solution for the ground-state; and they provide, self-consistently, the IAS energy via the charge-exchange Random Phase Approximation (RPA) \cite{Colo1994,Paar2004}. 
The results for the IAS energy, $E_{\rm IAS}$, as a function of 
$\Delta R_{\rm np}$ 
are plotted in Fig. \ref{fig1}. For the sake of completeness, results associated with other Skyrme interactions are also plotted in this figure. 

Results displayed in Fig. \ref{fig1} lie very close to a straight line. This can be understood as follows. The main contribution to $E_{\rm IAS}$ can be evaluated from the Coulomb direct contribution to the so-called displacement energy $\Delta E_d^{\rm C, direct}$ (cf. Ref. \cite{auerbach1969}). The latter quantity can be approximately calculated by assuming two uniform neutron and proton distributions of radius $R_n$ and $R_p$, respectively, and approximating the electric charge density with the proton density. This leads to $E_{\rm IAS}\approx \Delta E_d^{\rm C, direct} \approx \frac{6}{5}\frac{Ze^2}{r_0A^{1/3}}\left(1-\frac{1}{2}\frac{N}{N-Z}\frac{\Delta R_{\rm np}}{r_0A^{1/3}}\right)$ 
where 
$r_0 = \left( \frac{3}{4\pi\rho_0} \right)^{1/3}$ and thus
$2 r_0$ is the average distance between two nucleons in symmetric nuclear matter at saturation density. 
For the case of ${}^{208}$Pb, this formula predicts $E_{\rm IAS}\approx 20.9 -4.4 \Delta R_{\rm np}$ [MeV]. This result is in qualitative agreement with the linear fit to the microscopic calculations shown in Fig. \ref{fig1}, which gives $E_{\rm IAS} = 19.19(8)-5.0(2) \Delta R_{\rm np}$ [MeV], with a large correlation coefficient $r=-0.985$. The experimental IAS energy \cite{martin2007} is marked in the figure, 
and a simple extrapolation of the straight line would imply $\Delta R_{\rm np} = 0.07(2)$ fm. This value is incompatible with many independent analyses \cite{Lattimer:2013,Horowitz:2014,vinas14}. 
As a reference, recent experimental constraints from polarized proton elastic scattering \cite{zenihiro2010}, parity violating elastic electron scattering \cite{horowitz2012}, and electric dipole polarizability \cite{roca-maza2015}, are indicated in the bottom part of Fig. \ref{fig1}. 

To solve this puzzle, we have reconsidered all possible contributions to the IAS energy that have {\em not} been considered with sufficient care in self-consistent calculations so far. All the effects listed below are introduced in the HF mean-field calculations, as one can easily verify that they do not impact the proton-neutron RPA residual force. 
\begin{enumerate}
\item {\it Coulomb interaction: exact direct and exchange contributions.} The Coulomb energy per particle
is by far the dominant contribution to the IAS energy. Self-consistent RPA calculations of the IAS are exact, in the specific sense that they preserve the isospin symmetry: if the Coulomb effects are switched completely off, the IAS is found at zero energy \cite{lemmer}. When Skyrme forces are used, it is customary in the literature to adopt the Slater approximation for the Coulomb exchange. In the present case, we have instead included the exact Coulomb exchange. The detailed procedure is explained in Ref. \cite{AGDR}, where the reader can also find a detailed analysis for the case of $^{208}$Pb. The IAS energy is pushed upwards by $\approx$ 100 keV if exact Coulomb exchange is taken into account.   
\item {\em Electromagnetic spin-orbit contribution.}
The e.m. spin-orbit effect on the single-nucleon energy 
$\varepsilon_i$ is well known and reads \cite{isospin_book}
\begin{equation}\label{eq:emso_corr}
\Delta\varepsilon_i=\frac{\hbar^2c^2}{2m^2c^4}
x_i \langle\vec{l}_i\cdot\vec{s}_i\rangle
\int \frac{dr}{r}\frac{dU_{\rm Coul}}{dr}u_i^2(r),
\end{equation}
where $u_i(r)$ is the radial wave function, and $x_i$ is equal to $g_p-1$ 
in the case of protons and to $g_n$ in the case of neutrons 
($g_n=-3.82608545(90)$ and $g_p=5.585694702(17)$ are the neutron and 
proton $g$-factors, respectively \cite{codata2014}). The correction 
(\ref{eq:emso_corr}) can be estimated to be of the order of tens of keV,
and it is basically model-independent.
\item {\em Finite size effects} 
In most of the previous calculations, the Coulomb potential has been
evaluated by taking simply into account the point proton density, 
and identifying it with the charge density. 
In the present work, we
have considered in detail all contributions to the charge density which, 
when written in momentum space up to order $1/m^2$, 
reads \cite{ray1979}
\begin{eqnarray}
\rho_{\rm ch}(q)&=& \left(1-\frac{q^2}{8m^2}\right)
\left[G_{\rm E,p}(q^2)\rho_p(q) + 
G_{\rm E,n}(q^2)\rho_n(q)\right]\nonumber\\
&-&\frac{\pi q^2}{2m^2}\sum_{n,l,j,t} 
\left[2G_{\rm M,t}(q^2)-G_{\rm E,t}(q^2)
\right]\langle \vec{l}\cdot\vec{s}\rangle\times\nonumber\\
&&\int_0^\infty dr\frac{j_1(qr)}{qr}\vert u_{n,l,j}(r)\vert^2,
\end{eqnarray}
where $G_{\rm E,M}$ are the electromagnetic form factors, taken 
from \cite{friedrich2003}, $t$ labels either protons or neutrons 
and the sum runs over the occupied states. 
Finite-size effects spread out the Coulomb potential: its 
effects on the p-h excitations that make up the IAS are slightly 
weaker. The IAS energy decreases, albeit only by few tens of
keV, due to this effect. 
\item {\em Vacuum polarization correction}
From QED, the lowest-order correction to the fine-structure constant or 
to the Coulomb potential is the 
vacuum polarization: it amounts to a virtual emission and absorption of an electron-positron pair, and produces 
an extra repulsion that is not completely negligible even at 
the present low-energy scale. 
The corresponding potential can be written as follows 
(cf. Refs. \cite{Uehling:1935,Schwinger:1949} and Ref. \cite{wayne1976} 
in the case of a finite-size particle):
\begin{equation}
V_{\rm vp}(\vec{r})=\frac{2}{3}\frac{\alpha e^2}{\pi}
\int d\vec{r}^\prime\frac{\rho(\vec{r}^\prime)}{\vert\vec{r}-\vec{r}^\prime
\vert}\mathcal{K}_1\left(\frac{2}{\lambdabar_e}\vert\vec{r}-
\vec{r}^\prime\vert\right),
\end{equation}
where $\alpha$ the fine-structure constrant, $\lambdabar_e$ the 
reduced Compton electron wavelength, and
\begin{equation}
\mathcal{K}_1(x)\equiv\int_1^\infty dt e^{-xt}\left(\frac{1}{t^2}
+\frac{1}{2t^4}\right)\sqrt{t^2-1}.
\end{equation}
\item {\em Charge-symmetry breaking (CSB)  and charge-independence breaking (CIB)
forces.}
\end{enumerate}

\begin{table}[b!]
\vspace{-0.5cm}
\begin{center}
\caption{Effect of the different contributions from isospin breaking (including both CSB and CIB) mentioned in the text on the IAS energy in ${}^{208}$Pb. Corrections are basically model-indpendent except the last one.}
\begin{tabular}{lrr}
\hline\hline
& $E_{\rm IAS}$ [MeV]& Correction [keV]\\
\hline
No corrections
                                 & 18.31 &    \\
Exact Coulomb exchange           & 18.41 & +100\\
n/p mass difference              & 18.44 & +30\\
Electromagnetic spin-orbit       & 18.45 & +10\\
Finite size effects              & 18.40 & -50\\
Vacuum polarization ($V_{\rm ch}$) & 18.53 & +130\\
\hline
Isospin symmetry breaking        & 18.80 & +270\\
\hline\hline
\end{tabular}
\label{table:effects}
\end{center}
\end{table}

The contributions 1-4, together with the n-p mass difference which is one of the CSB potential terms, produce an overall upward shift of the IAS energy and 
the same for the straight line that connects the points of Fig. \ref{fig1}. 
In Table \ref{table:effects} we can see that this shift, by adding also the small effect of the neutron-proton mass difference, amounts to $\approx$ 220 keV and, consequently, it is 
too small in order to let the line intersect the experimental value for the IAS energy at a point that corresponds to a realistic range of $\Delta R_{\rm np}$ (indicated by the horizontal bars in Fig. 1). 

CSB and CIB effects have been widely discussed in the literature (see, for example,  Refs. \cite{HM79,Miller1990,Kolck1996}); however, most of the efforts have been devoted to study their impact on few-nucleon systems and 
few-hadron systems, or to derive them from QCD through Effective Field Theory (EFT) methods. 
Recently, the isospin mixing in $^{8}$Be was studied based on the Green's function Monte Carlo method by including the CSB interaction \cite{Wringa2013}. Although it is known for many years that CSB-CIB 
forces must be taken into account to reproduce the so-called Nolen-Schiffer anomaly along the
nuclear chart, the information on CSB-CIB forces in the nuclear medium is scarce. 
The nuclear shell model has been employed for quite some time to analyze the energies along 
the isobaric multiplets; recently, long sequences of multiplets in rotational bands have been used 
to determine the magnitude of CSB-CIB effects \cite{Bentley2007}. In the same context, it has been noticed that CSB-CIB interactions needed to explain the data are not 
consistent with those in the vacuum \cite{Bentley2015}. Similar conclusions
have been drawn in Ref. \cite{Baczyk2018}.

Therefore, 
in the present work we have kept our description simple and aimed to reconcile the scarce information about CSB-CIB in the medium, and the reproduction of the IAS energy, with a realistic value for the neutron skin. Borrowing from Ref. \cite{sagawa1995} [cf. Eqs. (18-21)], we define simpler 
Skyrme-like CSB and CIB interactions as follows,
\[
V_{\rm CSB}(\vec{r}_1,\vec{r}_2) \equiv 
\frac{1}{4}\left[\tau_z(1)+\tau_z(2)\right]
s_0(1+y_0P_\sigma)\delta(\vec{r}_1-\vec{r}_2),
\label{vcsb}
\]
and
\[
V_{\rm CIB}(\vec{r}_1,\vec{r}_2) \equiv \frac{1}{2}
\tau_z(1)\tau_z(2)
u_0(1+z_0P_\sigma)\delta(\vec{r}_1-\vec{r}_2). 
\label{vcib}
\]

$P_{\sigma}$ is the spin-exchange operator and we take  
$y_0=-1$ and $z_0=-1$.
The momentum-dependent terms written in Ref. \cite{sagawa1995} have not been
considered, under the rationale that the information that we have at
our disposal is not sufficient to pin down the values of all parameters of 
a general interaction with several partial waves.

We have looked at the ISB contributions to the energy per particle of symmetric nuclear matter, as predicted by the Brueckner-Hartree-Fock calculations of \cite{muther99} (based on AV18 \cite{av18}). We have determined a new Skyrme parameter set named SAMi-ISB, using the same fitting protocol of SAMi but including CSB-CIB contributions. 
Details are provided in the supplemental material. 
We have first started from existing parameters of SAMi and fixed the values of $s_0$ and $u_0$ by requiring a reproduction of the results of Ref. \cite{muther99} and the value of the IAS energy in ${}^{208}$Pb. This gives\footnote{The adopted deviations that lead to the estimated statistical errors are 30 keV on the IAS energy and 10 keV ($\sim$10\% error) on the CSB contribution to E/A in symmetric matter.} $s_0=-26.3(7)$ MeV fm$^3$ and $u_0=25.8(4)$ MeV fm$^3$. Then, the standard Skyrme parameters have been refitted; the effect of CSB/CIB is included but these forces affect the binding energies and charge radii only by a few per mil or per cent, so that this two-step strategy is feasible. 

\begin{table}[t!]
\vspace{-0.5cm}
\begin{center}
\caption{SAMi-ISB parameter set. 
The statistical errors $\sigma$ are given in parenthesis. 
See text for details.}
\begin{tabular}{lrlclrl}
\hline\hline
            &value($\sigma$)  &     & &   & value($\sigma$) &      \\
\hline
 $t_0$      &$-$2098.3(149.3)&MeV fm${}^{3}$   & &  $x_0$      &    0.242(9)&      \\
 $t_1$      &  394.7(15.8)&MeV fm${}^{5}$     & &  $x_1$      &   $-$0.17(33)&     \\
 $t_2$      & $-$136.4(10.8)&MeV fm${}^{5}$   & &  $x_2$      &   $-$0.470(4)&   \\
 $t_3$      &11995(686)&MeV fm${}^{3+3\alpha}$ & &  $x_3$      &    0.32(21)&     \\
 $W_0$      &  294(6)&                    & & & & \\ 
 $W_0^\prime$& $-$367(12)&                  & &  $s_0$ &$-26.3(7)$ &MeV fm$^3$  \\
 $\alpha$  &    0.223(31)&                  & & $u_0$ &$25.8(4)$ &MeV fm$^3$   \\
\hline\hline
\end{tabular}
\label{par}
\end{center}
\end{table}

The values of the SAMi-ISB parameters are provided in Table \ref{par}. 
As seen in Fig. \ref{fig1}, with SAMi-ISB, the IAS energy of 
$^{208}$Pb is predicted at $E_{\rm IAS}$ = 18.80(5) MeV 
($E^{\rm exp}_{\rm IAS} = 18.826\pm 0.010$ MeV \cite{martin2007}) 
with the neutron skin $\Delta R_{\rm np}$= 0.151(7) fm, which is within 
the realistic range deduced from the three experiments. 
The quality of this interaction is similar or better than SAMi if we look at overall properties like those in uniform matter.
The values of $J$ and $L$ are, in the case of SAMi-ISB [SAMi], 30.8(4) 
MeV [28(1) MeV] and 50(4) MeV [44(7) MeV]. These are quite realistic
values, according to our general discussion at the start of this 
paper. While the detailed assessment of SAMi-ISB along the isotope chart
is out of our scope here, we show in Table \ref{resfit} some results 
for binding energies, charge radii, and neutron skin
thicknesses. Moreover, we have checked the predictive power of SAMi-ISB by
calculating the IAS energy in other nuclei. In the Sn isotopes with A=112-124, 
the IAS energies calculated with SAMi differ from the experimental values by 
about 600 keV, while this discrepancy is reduced to $\approx$ 50-200 keV by using 
SAMi-ISB. In $^{40}$Ca and $^{90}$Zr the results obtained with SAMi-ISB are also
improved with respect to SAMi (cf. the supplemental material for more information). 

\begin{table}[h!]
\vspace{-0.5cm}
\begin{center}
\caption{Experimental data 
\cite{ame16,angeli2013}
and SAMi-ISB results for the binding energies
($B$), charge radii ($r_c$), and neutron skin thickness ($\Delta R_{\rm np}$) for some selected nuclei.}
\begin{tabular}{lrrrrrrrr}
\hline\hline
El.& $N$ & $B\quad$ & $B^{\rm exp}$ & $r_{\rm c}$ & $r^{\rm exp}_{\rm c}$ & $\Delta R_{\rm np}$\\
   &     &[MeV]     & [MeV]  & [fm] & [fm] & [fm] \\
\hline
Ca &  28 &417.67& 415.99&3.49&3.47 &0.214\\
Zr &  50 &783.60& 783.89&4.26&4.27 &0.097\\
Sn &  82 &1102.75&1102.85&4.73& -- &0.217\\
Pb & 126 &1635.78&1636.43&5.50&5.50&0.151\\
\hline\hline
\end{tabular}
\label{resfit}
\end{center}
\end{table}

In conclusion, SAMi-ISB is a new parameterization of a Skyrme-like energy density functional (EDF) that reconcile standard nuclear properties (saturation density, binding energy and charge radii of finite nuclei) with {\em both} our current understanding of the density behavior of the symmetry energy {\em and} the reproduction of the IAS energy of $^{208}$Pb as well as in Sn isotopes.  
We have self-consistently included for the first time within the HF+RPA framework all known contributions that break the isospin symmetry. All of these contributions are calculated in a model-independent way, except the CSB-CIB contribution. 
We have fixed only two free parameters 
in the CSB-CIB terms, 
and we have shown that this allows reproducing at the same time BHF calculations based on AV18, and the IAS energy of a heavy nucleus such as ${}^{208}$Pb without compromising the other properties of nuclear matter and
finite nuclei.

Funding from the European Union's Horizon 2020 research and innovation
programme under grant agreement No 654002 is acknowledged.

\bibliography{bibliography}

\end{document}


%
\title{Supplemental material: The nuclear symmetry energy and the breaking of the \\ isospin symmetry: how do they reconcile with each other?}

\author{X. Roca-Maza}
\email{xavier.roca.maza@mi.infn.it}
\affiliation{Dipartimento di Fisica, Universit\`a degli Studi di Milano and INFN, Sezione di Milano, Via Celoria 16, 20133 Milano, Italy.}

\author{G.  Col\`o}
\email{colo@mi.infn.it}
\affiliation{Dipartimento di Fisica, Universit\`a degli Studi di Milano and INFN,  Sezione di Milano, Via Celoria 16, 20133 Milano, Italy.}

\author{H. Sagawa}
\email{hiroyuki.sagawa@gmail.com}
\affiliation{RIKEN Nishina Center, Wako 351-0198, Japan\\
             Center for Mathematics and Physics, University of Aizu, Aizu-Wakamatsu, Fukushima 965-8560, Japan}

\date{\today}

\begin{abstract}
This supplemental material contains some technical details on the Hartree-Fock plus Random Phase Approximation calculations reported in the manuscript and on the fitting protocol and covariance analysis of the newly introduced interaction SAMi-ISB. 
\end{abstract}

\pacs{}

\maketitle

\section{Hartree-Fock plus Random Phase Approximation based on the Skyrme effective interaction}
The calculations are done within the framework of the Skyrme Hartree-Fock (HF) plus charge-exchange Random Phase Approximation (RPA) \cite{Colo1994,AGDR}. We adopt the standard form of Skyrme interactions with the notations of Ref. \cite{chabanat1997}. Our charge-exchange RPA had been first introduced and illustrated in Ref. \cite{Colo1994} and further improved in Ref. \cite{AGDR}, to which we refer the reader for details. Some details that are common between charge-exchange and normal RPA can be found in Ref. \cite{colo2013}, where HF-RPA for non charge-exchange models is fully discussed. 

The technical details of the calculations are as follows. The ground state properties of the different nuclei are calculated in coordinate space using box boundary conditions. The radius of the box is taken to be 20 fm. The same box is used to calculate discrete states at positive energy that are associated with the continuum part of the spectrum. A cutoff energy of 60 MeV (in the single-particle energy) is adopted for the RPA calculations. With this energy cutoff, we have checked that the non-energy weighted sum rule for the Isobaric Analog State is satisfied. The numerical accuracy of the HF+RPA code in the calculation of the $E_{\rm IAS}$ is within 10 keV for the test case of ${}^{90}$Zr \cite{AGDR}.  

\section{Fitting protocol and covariance analysis}

{\it Fitting protocol.} The fitting protocol of SAMi-ISB is based on a $\chi^2$ minimization and it is identical to that reported in Ref. \cite{RocaMaza2012} --used to build the SAMi interaction-- except for the inclusion in the $\chi^2$ of: (i) the experimental value of the excitation energy of the Isobaric Analog State in ${}^{208}$Pb \cite{antony1997} ($E_{\rm IAS}$) with an adopted error of $\delta E_{\rm IAS}= 30$ keV in agreement with the very accurate determination of this quantity; and (ii) the isospin symmetry breaking (ISB) contributions to the symmetric nuclear matter equation of state (EoS) as calculated by the benchmark Brueckner-Hartree-Fock (BHF) calculations based on AV18 \cite{av18} reported in reference \cite{muther99}. We have fitted 9 data points (as shown in Fig. \ref{bhf}) and the adopted error for each of them has been taken to be 10 keV. We include the latter set of pseudo-data to test if one can reconcile the more fundamental BHF results based on the most complete Argonne potential with the experimental data on ${}^{208}$Pb on the grounds of a mean-field approach. The minimization procedure have been implemented via the {\sc MINUIT} algorithm \cite{minuit} as for the case of SAMi.

\begin{figure}[t!]
\includegraphics[width=\linewidth,clip=true]{csb-cib-AV18.eps}
\caption{(Color online) ISB contributions to the symmetric matter equation of state as predicted by the BHF calculations based on AV18 potential (black circles). The solid line is the optimal fit of this quantity.}
\label{bhf}
\end{figure}

{\it Parameter set.} Nowadays the nuclear community has a clear idea on the ranges of acceptable values for some nuclear matter properties around saturation (that is, Equation of State parameters). It is, therefore, customary and convenient to use these nuclear matter properties as fitting parameters \cite{kortelainen2010}. The relation between them and the original parameters of the Skyrme interaction can be found, e.g. in Refs. \cite{chabanat1997,kortelainen2010}. Specifically, the parameters used for the present fit are the saturation energy $e_0$, saturation density $\rho_0$, nuclear matter incompressibility $K_0$, symmetry energy at saturation $J$, symmetry energy slope parameter $L$, isoscalar effective mass $m^*_{\rm IS}$, isovector effective mass $m^*_{\rm IV}$ as well as the two spin-orbit parameters $W_0$ and $W_0^\prime$. In addition to these parameters, two extra parameters related to ISB in nuclei have been also added, as discussed in the manuscript.

For simplicity, the fitting procedure has been divided into two steps due to the fact that ISB effects on observables such as masses and charge radii used in the fitting procedure are reasonably small (around a few \% or even less). First, based on the SAMi interaction, ISB parameters have been determined to reproduce both $E_{\rm IAS}$ in ${}^{208}$Pb and the results of Ref. \cite{muther99}, shown here in Fig. \ref{bhf}. Subsequently, the second and last step has consisted in a fine tunning of the remaining parameters in order to keep the description of nuclear masses, charge radii, etc.,  at the level of accuracy of the original SAMi interaction and of other currently available models of the same type.

\begin{table}[t!]
\begin{center}
\caption{SAMi-ISB parameter set used in the fit. See text for details.}
\begin{tabular}{lrlclrl}
\hline\hline
 $e_0$    & -16.03(2)& MeV & & $L$            & 50(5)      & MeV\\
 $\rho_0$ & 0.1613(6)& MeV & & $m^*_{\rm IS}/m$ & 0.730(19) & \\
 $K_0$    &  235(4)  & MeV & & $m^*_{\rm IV}/m$ & 0.667(116) & \\ 
 $J$      &  30.8(4) & MeV & & $G_0$          & 0.15(fixed)&\\  
 $W_0$      &  294(6)      &  & & $G_0^\prime$    &0.35(fixed)&\\
$W_0^\prime$& $-$367(12)&  &   &               &             & \\
           &           &  &   &               &             & \\
$s_0$ &$-26.3(7)$ &MeV fm$^3$ & & $u_0$ &$25.8(4)$ &MeV fm$^3$ \\
\hline\hline
\end{tabular}
\label{par}
\end{center}
\end{table}

\begin{figure}[b!]
\includegraphics[width=\linewidth,clip=true]{correlations.eps}
\caption{Correlation coefficient matrix (in absolute value) for the SAMi-ISB parameters.}
\label{correlations}
\end{figure}

{\it Covariance analysis.} The covariance analysis has been made following the standard approximations reported in Ref. \cite{roca-maza2015} as well as in other works and textbooks.  

In Table \ref{par} the optimal value of the fitted parameters is given together with the associated statistical theoretical errors (within parenthesis). The relatively small errors reported here for some of the parameters when compared with other available parameterizations are a consequence of the information included in the fit via the $\chi^2$. That is, the fitted information constrains rather well all the channels (parameters) of the interaction. In comparison with other models of the same type, this has been possible due to the inclusion of pseudo-data: specially, the neutron matter equation of state in the range of densities $0.07$ fm$^{-3}<\rho<0.4$ fm$^{-3}$ and the empirically suggested values of the spin $G_0$ and spin-isospin $G_0^\prime$ Landau parameters at nuclear saturation density (see Ref. \cite{RocaMaza2012} for further details and discussions).   

\begin{table}[t!]
\begin{center}
\caption{SAMi-ISB in terms of Skyrme standard parameters. See text for details.}
\begin{tabular}{lrlclrl}
\hline\hline
            &value($\sigma$)  &     & &   & value($\sigma$) &      \\
\hline
 $t_0$      &$-$2098.3(149.3)&MeV fm${}^{3}$   & &  $x_0$      &    0.24(9)&      \\
 $t_1$      &  394.7(15.8)&MeV fm${}^{5}$     & &  $x_1$      &   $-$0.17(33)&     \\
 $t_2$      & $-$136.4(10.8)&MeV fm${}^{5}$   & &  $x_2$      &   $-$0.47(4)&   \\
 $t_3$      &11995(686)&MeV fm${}^{3+3\alpha}$ & &  $x_3$      &    0.32(21)&     \\
 $W_0$      &  294(6)&                    & & & & \\ 
 $W_0^\prime$& $-$367(12)&                  & &  $s_0$ &$-26.3(7)$ &MeV fm$^3$  \\
 $\alpha$  &    0.223(31)&                  & & $u_0$ &$25.8(4)$ &MeV fm$^3$   \\
\hline\hline
\end{tabular}
\label{parsky}
\end{center}
\end{table}

In Fig. \ref{correlations}, the correlations between the fitted parameters are shown in absolute value. The correlations are from moderate to very low. For the same reasons highlighted before, this is because the information included in the fit via the $\chi^2$ is constraining all the channels (parameters) of the interaction. As an example to understand this result, one can focus on the isovector channel of the effective interaction parametrized around saturation by $J$ and $L$: it is well known that information on binding energies and charge radii constraints the value of $e_0$ and $\rho_0$ in a tight and model independent way, and that it does not tightly constraint $J$ and $L$. The inclusion of the neutron matter equation of state in the fit, forces the value of $J$ once $e_0$ is fixed from information on binding energies and charge radii; and the fact of including the neutron matter equation of state at different densities constraints, by itself alone, the value of $L$. For a more detailed discussion on this, cf. Ref. \cite{roca-maza2015}.     

For completeness, we report in Table \ref{parsky} the values and associated statistical errors of the standard Skyrme parameters univoquely determined from the optimal fitted parameters reported in Table \ref{par}. 

\begin{table}[t!]
\begin{center}
\caption{Some relevant observabales for ${}^{208}$Pb with associated statistical errors ($\sigma$). See text for details.}
\begin{tabular}{lrrrr}
  \hline\hline
Observable        & Value    & $\sigma_{\rm IC}$& $\sigma_{\rm ISB}$&$\sigma$\\
\hline
$B$ [MeV]         & $-$1635.8& 1.3&   1.1   & 1.7    \\
$r_n$ [fm]        & 5.600    & 0.007& 0.0008& 0.007  \\
$r_p$ [fm]        & 5.449    &0.002&  0.0003& 0.002  \\
$r_{\rm ch}$ [fm]   & 5.507   &0.002& 0.0003 & 0.002   \\
$E_{\rm IAS}$ [MeV] & 18.80   &0.05&  0.02   &   0.05  \\
\hline\hline
\end{tabular}
\label{obs}
\end{center}
\end{table}

The covariance analysis, allows also to estimate the associated errors on predicted observables. This is of relevance in order to assess (at least partially) the reliability of the model when doing extrapolations. Since we focus on the $E_{\rm IAS}$ in ${}^{208}$Pb, we give as example in Table \ref{obs} the value for different observables and associated errors. In this Table, the error budget has been divided according to the two step minimization procedure where $\sigma_{\rm IC}$ stands for the error coming from the isospin-conserving terms of the effective interaction and $\sigma_{\rm ISB}$ from the ISB terms. The final error is just $\sigma=\sqrt{\sigma_{\rm IC}^2+\sigma_{\rm ISB}^2}$.

\begin{table}[b!]
\begin{center}
\caption{$E_{\rm IAS}$ and $r_{\rm ch}$ for some selected nuclei as predicted by SAMi and SAMi-ISB, as well as the experimental values and errors (within parenthesis).}
\begin{tabular}{|lr|rr|rr|rr|}
\hline
&     &\multicolumn{2}{c}{SAMi}  &\multicolumn{2}{c}{SAMi-ISB}&\multicolumn{2}{c|}{Experiment}\\
\hline
El.& $N$ & $E_{\rm IAS}$ & $r_{\rm ch}$ & $E_{\rm IAS}$ & $r_{\rm ch}$ & $E_{\rm IAS}$ & $r_{\rm ch}$ \\
   &     &[MeV] & [fm] & [MeV] & [fm] &  [MeV] & [fm]  \\
\hline
Ca & 28 & 6.573&3.525& 6.79(2)&3.497(3)& 7.182(8~) &3.477(2)\\
Zr & 50 &11.199&4.283&11.36(4)&4.262(3)&11.901(12)&4.269(1)\\
Sn &112 &13.408&4.539&14.27(3)&4.611(3)&14.019(20)&4.595(2)\\
   &114 &13.347&4.553&14.17(3)&4.619(3)&13.940(20)&4.610(2)\\           
   &116 &13.288&4.567&14.06(3)&4.633(3)&13.861(20)&4.625(2)\\           
   &118 &13.220&4.582&13.97(3)&4.647(3)&13.728(17)&4.639(2)\\           
   &120 &13.158&4.596&13.91(3)&4.659(2)&13.667(32)&4.652(2)\\           
   &122 &13.090&4.610&13.76(3)&4.671(2)&13.667(20)&4.663(2)\\           
   &124 &13.027&4.623&13.66(3)&4.684(2)&13.596(20)&4.673(2)\\           
Pb &126 &18.256&5.517&18.80(5)&5.507(2)&18.826(10)&5.501(1)\\
\hline
\end{tabular}
\label{pred}
\end{center}
\end{table}

\section{SAMi-ISB predictions}


\begin{figure}[t!]
\vspace{5mm}  
\includegraphics[width=\linewidth,clip=true]{sn-ias.eps}
\caption{$E_{\rm IAS}$ for Sn isotopes as predicted by SAMi and SAMi-ISB, compared to experimental data.}
\label{snias}
\end{figure}

In Table \ref{pred} and Fig. \ref{snias}, the values of $E_{\rm IAS}$ for the case of some measured Sn isotopes, as predicted by SAMi and SAMi-ISB, are shown. As expected, the results have been remarkably improved with respect to the results of the SAMi interaction. The IAS energies calculated with SAMi differ from the experimental values by about 600 keV, while this discrepancy is reduced
to $\approx$ 50-200 keV by using SAMi-ISB. In these calculations pairing correlations have been neglected since we have checked that their effects are of the order of the statistical error reported here. From Fig. \ref{snias}, it is clear that the isotopic trends are well reproduced by both interactions. In Table \ref{pred}, we extend our predictions also to ${}^{48}$Ca and ${}^{90}$Zr. In these cases, the results are also improved with respect to SAMi. Specifically, SAMi-ISB corrects by about 200 keV the values predicted by SAMi.

\bibliography{bibliography2}